\documentclass[aoas,MSNbibl,nameyear,dvips]{arximspdf}
\usepackage{graphicx}
\doi{10.1214/12-AOAS607} 
\volume{7}
\issue{2}
\pubyear{2013}
\firstpage{799}
\lastpage{822}

\makeatletter

\renewcommand{\mid}{|}

\def\regtm{$^{\bigcirc\hspace{-6pt} \texttt{\tiny R}}$}

\makeatother

\begin{document}
\begin{frontmatter}

\title{Sparse latent factor models with interactions: Analysis of gene
expression data\thanksref{T0}}
\runtitle{Sparse latent factor models with interactions}

\thankstext{T0}{Supported by Award Number U54CA112952 from the National
Cancer Institute, by funding from the Measurement to Understand
Re-Classification of Disease of Cabarrus and Kannapolis \mbox{(MURDOCK)} Study
and by a grant from the NIH CTSA (Clinical and Translational Science
Award) 1UL1RR024128-01 to Duke University.}

\begin{aug}
\author[A]{\fnms{Vinicius Diniz} \snm{Mayrink}\corref{}\thanksref{t1}\ead[label=e1]{vdm@est.ufmg.br}}
\and
\author[B]{\fnms{Joseph Edward} \snm{Lucas}\ead[label=e2]{joseph.lucas@duke.edu}}
\runauthor{V. D. Mayrink and J. E. Lucas}
\affiliation{Universidade Federal de Minas Gerais and Duke University}
\address[A]{Departamento de Estatistica, ICEx\\
Universidade Federal de Minas Gerais \\
Av Antonio Carlos, 6627, Pampulha \\
Belo Horizonte, MG, 31270-901 \\
Brazil\\
\printead{e1}} 
\address[B]{Institute for Genome Sciences\\
\quad and Policy \\
Duke University\\
CIEMAS, Box 3382 \\
Durham, North Carolina 27708 \\
USA\\
\printead{e2}}
\end{aug}

\thankstext{t1}{This study was developed in the Ph.D. program of the
Department of Statistical Science at Duke University where the first
author received his Ph.D. degree.}

\received{\smonth{3} \syear{2012}}
\revised{\smonth{8} \syear{2012}}

%
\begin{abstract}
Sparse latent multi-factor models have been used in many exploratory
and predictive problems with high-dimensional multivariate
observations. Because of concerns with identifiability, the latent
factors are almost always assumed to be linearly related to measured
feature variables. Here we explore the analysis of multi-factor models
with different structures of interactions between latent factors,
including multiplicative effects as well as a more general framework
for nonlinear interactions introduced via the Gaussian Process. We
utilize sparsity priors to test whether the factors and interaction
terms have significant effect. The performance of the models is
evaluated through simulated and real data applications in genomics.
Variation in the number of copies of regions of the genome is a
well-known and important feature of most cancers. We examine
interactions between factors directly associated with different
chromosomal regions detected with copy number alteration in breast
cancer data. In this context, significant interaction effects for
specific genes suggest synergies between duplications and deletions in
different regions of the chromosome.
\end{abstract}

%

\end{frontmatter}

\section{Introduction}\label{sec1}

In recent years, numerous studies have applied factor models combined
with the Bayesian framework to analyze gene expression data, and their
results often show an improvement in the identification and estimation
of metagene groups and patterns related to the underlying biology; see,
for example, \citet{West2003}, \citet{LucasEtAl2006} and
\citet
{CarvalhoEtAl2008}. The usual formulation for factor models assumes
additive effects of latent factors across the samples. This assumption
leads to very tractable model fitting and computation, but may not
represent the reality in some applications. The structure of dependence
between genes in biological pathways motivates the idea of a model with
nonlinear interactions between latent factors. The presence of
interactions can have important implications for the interpretation of
the underlying biology.

The study of nonlinear interactions between observed variables has been
the focus of many publications in the context of regression problems.
In many cases, the proposed model introduces the nonlinearity through
the specification of Gaussian Process (GP) priors. \citet
{HenaoWinther2010} consider sparse and identifiable linear latent
variable (factor) and linear Bayesian network models for parsimonious
analysis of multivariate data. The framework consists of a fully
Bayesian hierarchy for sparse models using spike and slab priors,
non-Gaussian latent factors and a stochastic search over the ordering
of the variables. The authors argue that the model is flexible in the
sense that it can be extended by only changing the prior distribution
of a set of latent variables to allow for nonlinearities between
observed variables through GP priors.

The nonlinear relationship between a set of observed variables is also
the topic of \citet{HoyerEtAl2009} in the context of Directed Acyclic
Graphs (DAG). Each observed variable (node in a DAG) is obtained as a
function of its parents plus independent additive noise. An arbitrary
function is chosen to define linear/nonlinear relationships between the
observed values. The paper evaluates whether a DAG is consistent with
the data by constructing a nonlinear regression of each variable on its
parents, and subsequently testing whether the resulting residuals are
mutually independent. GP regression and kernelized independence tests
are used in the paper.

Associations between observed and latent variables is another
interesting topic. \citet{ArmingerMuthen1998} consider latent variable
models including polynomial terms and interactions of latent regressor
variables. Two groups of observed variables are used: the response
vector \textbf{y}, and the vector of covariates \textbf{x}. Their model
specifies two equations; the first one expresses \textbf{y} as a linear
combination of polynomial terms and/or interactions of elements in the
latent vector $\xi$. The second equation defines a factor model without
interaction terms, where $\xi$ is the factor score and \textbf{x} is
the target data. Because the model includes components representing
functions of latent variables in the first equation, the authors denote
the formulation as nonlinear. They use the Bayesian framework with
conjugate priors to estimate the parameters; sparsity priors are not
considered in their analysis. In the spirit of factor analysis,
\citet
{TehEtAl2005} model the relationships among components of a response
vector \textbf{y} using linear (or generalized linear) mixing of
underlying latent variables indexed by a covariate vector \textbf{x}
(observed values). The authors assume that each latent variable is
conditionally independently distributed according to a GP, with \textbf
{x} being the (common) index set. The mean of the response \textbf{y}
is then a function of a linear combination of the conditionally
independent GP.

Most applications of GP models involve learning tasks where both output
and input data are assumed to be given at training time. \citet
{Lawrence2004} and \citet{Lawrence2005} have proposed a
multiple-output GP regression model assuming observed output data and
latent variables as inputs. The approach explores nonlinear
interactions between the latent factors. The authors introduce a
probabilistic interpretation of principal component analysis (PCA)
named dual probabilistic PCA (DPPCA). The DPPCA model has the advantage
that the linear mappings from the latent-space to the data-space can be
easily nonlinearized through Gaussian processes (DPPCA with a GP
introducing nonlinearity is then called GP Latent Variable Model or
GP-LVM). The GP (assumed for latent variables) with an inner product
kernel in the covariance function defines a linear association, and it
has an interpretation as a probabilistic PCA model. GP-LVM can be
obtained by replacing this inner product kernel with a nonlinear
covariance function. The nonlinear mappings are designed to address the
weaknesses in visualizing data sets that arise when using statistical
tools that rely on linear mappings, such as PCA and standard factor
models. The analyses are based on optimization via maximum likelihood
estimation; no MCMC algorithm is applied and no sparsity prior is assumed.

In GP models, inference is analytically tractable for regression
problems, and deterministic approximate inference algorithms are widely
used for classification problems. The use of MCMC methods to sample
from posterior distributions in a model assuming GP prior has been
explored in the literature only for cases with observed input data. As
an example, \citet{TitsiasEtAl2009} describe an MCMC algorithm which
constructs proposal distributions by utilizing the GP prior. At each
iteration, the algorithm generates control variables and samples the
target function from the conditional GP prior. The control variables
are auxiliary points associated with observed input variables defined
in the model. An advantage of MCMC over deterministic approximate
inference is that the sampling scheme will often not depend on details
of the likelihood function, and is therefore very generally applicable.
In addition, the development of deterministic approximations is
difficult since the likelihood can be highly complex. \citet
{ChenEtAl2010} have considered inference based on Variational
Bayesian (VB) approximation and Gibbs sampling to examined distinct
ways of inferring the number of factors in factor models applied to
gene expression data. The study indicates that while the cost of each
VB iteration is larger than that of MCMC, the total number of VB
iterations is much smaller. However, the CPU cost of MCMC appears to be
worthwhile, as they found that for a large-scale data set the MCMC
results were significantly more reliable than VB; the VB approximation
has difficulties with local-optimal solutions, and the factorized form
of the VB posterior may not be as accurate for large-scale problems.

Different latent class models have been proposed in the literature to
analyze the DNA Copy Number Alteration (CNA) problem. For example,
\citet
{LucasEtAl2010} use sparse latent factor analysis to identify CNA
associated with the hypoxia and lactic acidosis response in human
cancers. Specifically, they fit a latent factor model of the gene
signatures in one data set of 251 breast tumors [\citet
{MillerEtAl2005}] to generate 56 latent factors. These factors then
allow for connections to be made between a number of different data
sets, which can be used to generate biological hypotheses regarding the
basis for the variation in the gene signatures. They have identified
variation in the expression of several factors which are highly
associated with CNAs in similar or distinct chromosomal regions.
\citet
{DeSantisEtAl2009} developed a supervised Bayesian latent class
approach to evaluate CNA on array CGH data. The authors assume that
tumors arise from subpopulations (latent classes) sharing similar
patterns of alteration across the genome. The methodology relies on a
Hidden Markov Model (HMM) to account for the dependence structure
involving neighboring clones within each latent class. In particular,
the approach provides posterior distributions that are used to make
inferences about gains and losses in copy number. \citet
{FridlyandEtAl2004} proposed a discrete-state homogeneous HMM where
underlying states are considered segments of a common mean. One of the
goals of the procedure is to identify copy number transitions.
\citet
{MarioniEtAl2006} extended this approach by developing the method
BioHMM for segmenting array CGH data into states with the same
underlying copy number. They use a heterogeneous HMM with probability
of transitioning between states depending on the distance between
adjacent clones.

We are interested in the study of multi-factor models developed for the
analysis of matrices representing gene expression patterns. Our goal is
to investigate the existence of interaction effects involving latent
factors. In order to test the significance of the interaction terms,
the mixture prior with a point mass at zero and a Gaussian component
(sometimes referred to as the ``spike and slab'' prior) is assumed. This
type of prior has been used effectively to define the sparse structure
in \citet{West2003}, \citet{LucasEtAl2006}, \citet
{CarvalhoEtAl2008}
and others. The outline of this paper is as follows. In Section \ref{sec2} we
propose a factor model with multiplicative interactions between latent
factors. Our approach for this problem has not yet been considered in
the literature. Two strategies are used to introduce the interactions.
Section \ref{sec3} explores nonlinear structure of interactions between factors;
the formulation is more general. In short, we introduce nonlinearities
through the specification of a GP prior for a set of latent variables.
Five different versions of the model are investigated; they differ in
terms of prior formulations for probability parameters and the
assumption regarding the similarity of the interaction effects for
distinct features. In Section \ref{sec4} a simulated study is developed to
evaluate and compare the models from Sections \ref{sec2} and \ref{sec3}. Additional
synthetic data analyses to assess the performance of the models are
presented in \citet{MayrinkLucas2012}. Sections \ref{sec5} and \ref{sec6} show real data
applications where we examine interaction effects related to
chromosomal regions detected with CNA in breast cancer data. Finally,
Section \ref{sec7} indicates the main conclusions and future work.

The algorithms required to fit the proposed models are implemented
using the MATLAB programming language (\url{http://www.mathworks.com}).

\section{Factor model with multiplicative interactions}\label{sec2}

Assume $X$ is an ($m \times n$) matrix with $X_{ij}$ representing gene
$i$ and sample $j$. We propose the model
%
\begin{equation}\label{Eq01}
X = \alpha\lambda+ \theta\eta+ \varepsilon,
\end{equation}
where $\alpha$ is an ($m \times L$) matrix of loadings,
$\lambda$ is an ($L \times n$) matrix of factor scores, $\theta$ is an
($m \times T$) matrix of loadings, $\eta$ is a ($T \times n$) matrix of
interaction effects, and $\varepsilon$ is an ($m \times n$) noise matrix
with $\varepsilon_{ij} \sim N(0,\sigma_i^2)$; let
$\sigma^2 = (\sigma_1^2,\ldots,\sigma_m^2)'$. With this formulation,
we are separating the linear and nonlinear effects. One could add the
term $\mu1_n$ in this model to estimate the mean expression of the
genes; $\mu$ is an $m$-dimensional column vector and $1_n$ is an
$n$-dimensional row vector of ones. We prefer the parsimonious version
where the rows of $X$ are standardized and $\mu= \mathbf{0}$
is assumed.

The multiplicative interactions are defined in $\eta$ with the
following assumption: $\eta_{1j} = \lambda_{1j} \lambda_{2j}$,
$\eta_{2j} = \lambda_{1j} \lambda_{3j},\ldots,\eta_{Tj} = \lambda
_{(L-1)j} \lambda_{Lj}$. Note that $T = L!/[(L-2)! 2!]$.

In terms of prior distributions, we consider the conjugate
specifications $\lambda_{lj} \sim N(0,1)$ and $\sigma_i^2 \sim
\mathit{IG}(a,b)$. In our study, the bimodal sparsity promoting priors are key
elements in the structure of the model. This form of prior originated
in the context of Bayesian variable selection, and it has been the
subject of substantial research; see George and
McCulloch (\citeyear{GeorgeMcCulloch1993,GeorgeMcCulloch1997})
and \citet{Geweke1996}. The spike and slab
mixture prior is defined for the factor loadings to allow for sparsity
and to test whether the factors/interactions have significant effect on
each gene. Assume
%
\begin{eqnarray}
\label{Eq02}
\alpha_{il} &\sim& (1-h_{il}) \delta_0(
\alpha_{il}) + h_{il} N(0,\omega_{\alpha}),
\nonumber\\[-8pt]\\[-8pt]
h_{il} &\sim& \operatorname{Bernoulli}(q_{il}) \quad\mbox{and}\quad
q_{il} \sim\operatorname{Beta}(\gamma_1,\gamma_2),
\nonumber
\\
\label{Eq03}
\theta_{it} &\sim& (1-z_{it}) \delta_0(
\theta_{it}) + z_{it} N(0,\omega_{\theta}),
\nonumber\\[-8pt]\\[-8pt]
z_{it} &\sim& \operatorname{Bernoulli}(\rho_{it}) \quad\mbox{and}\quad
\rho_{it} \sim\operatorname{Beta}(\beta_1,\beta_2).\nonumber
\end{eqnarray}
We consider two approaches to introduce the corresponding
multiplicative interaction term; they are enumerated below:
\begin{longlist}[(2)]
\item[(1)] Introduce the interaction via Gaussian prior: $\eta_{tj} \sim
N(\lambda_{l_1 j} \lambda_{l_2 j}, \nu)$.
\item[(2)] Assume the product with probability 1: $\eta_{tj} = \lambda_{l_1
j} \lambda_{l_2 j}$.
\end{longlist}
In the cases above, let $l_1 < l_2 \in\{1,\ldots,L\}$ be the
indices of factors involved in the product term related to $\eta_{tj}$
where $t \in\{1,\ldots,T\}$.

In the first version, we specify the product $\lambda_{l_1 j}
\lambda_{l_2 j}$ as the mean parameter of the Gaussian distribution.
This approach can be generalized with the specification of any function
$f(\lambda_{l_1 j},\lambda_{l_2 j})$, which makes it possible
to investigate other types of relationships between factors. The
variance $\nu$ must have a small value; otherwise, we are indicating a
weak association between $\eta_{tj}$ and $\lambda_{l_1 j}
\lambda_{l_2 j}$. In this case, the multiplicative effect is lost and
the interaction factor is just another factor in the model. If the
number of genes is large, the variability in the posterior distribution
can be very small due to the large amount of data. In this case, $\nu$
is difficult to set and only extremely small values will ensure that
$\eta_{tj}$ is associated with $\lambda_{l_1 j} \lambda_{l_2
j}$. The target posterior in approach 1 is
$p(\alpha,\lambda,\theta,\eta,\sigma^2|X)$.

In the second approach, we force the perfect association between the
interaction factor and the corresponding product term; this strategy is
convenient to deal with large data sets. Here,
$p(\alpha,\lambda,\theta,\sigma^2|X)$ is the target posterior
distribution. Note that $\eta_{tj}$ is regarded as fixed variables;
$\eta_{tj} = \lambda_{l_1 j} \lambda_{l_2 j}$.

A Gibbs Sampler algorithm is implemented to generate observations from
the target posterior distributions; see Section A in \citet
{MayrinkLucas2012} to identify the full conditional distributions. A
simulated study has been developed to investigate the performance of
the proposed model; Section B in \citet{MayrinkLucas2012} shows the
results and the associated discussion.

\section{Factor model with general nonlinear interactions}\label{sec3}

Assume the model
%
\begin{equation}\label{Eq04}
X = \alpha\lambda+ F + \varepsilon,
\end{equation}
where $\alpha$ is an ($m \times L$) matrix of loadings,
$\lambda$ is an ($L \times n$) matrix of factor scores, and $\varepsilon$
is an ($m \times n$) matrix with idiosyncratic noise terms
$\varepsilon_{ij} \sim N(0,\sigma_i^2)$. Here, we replace the term
$\theta\eta$ with $F$, which is an ($m \times n$) matrix of
interaction effects. This model is defined with $L$ factors, $m$
features and $n$ samples. Again, we chose to work without the genes'
mean expression parameter $\mu$. This parsimonious configuration
reduces the computational cost to fit large real data sets. In all
applications, the rows of $X$ are standardized to define $\mu=
\mathbf{0}$.

If no constraint is imposed to $\alpha\lambda$ and $F$, the model will
experience identifiability issues. As an example, consider the $i$th
row of $\alpha\lambda+ F$ and note that $\alpha_{i \cdot}
\lambda+ F_{i \cdot} = C \alpha_{i \cdot} \lambda+ F^*_{i \cdot}$,
where $F^*_{i \cdot} = (1-C) \alpha_{i \cdot} \lambda+ F_{i
\cdot}$ and $C$ is any real number. This paper is focused on the
analysis of gene expression data; however, one should not restrict the
application to this context only. The methodology can be applied to any
data set satisfying the following aspects: (i) the data matrix $X$
can be specified with rows${}={}$features/variables and columns${}={}$samples,
(ii) at least two factors can be well defined, (iii) for each
factor ``$l$'' there is a subset of features $G_l$ in $X$ which are
linearly related to that factor with no interaction effects. Our goal
is to identify interactions between factors and identify the features
in the data that are affected by such interactions. We take advantage
of the known feature--factor relationship involving the elements in
$G_l$ to impose, via prior distributions, a specific configuration for
$\alpha$ and $F$ in (\ref{Eq04}); see Section D in
\citet{MayrinkLucas2012}. In particular, we assume that most features
are not affected by interactions; therefore, prior distributions
favoring $F_{i \cdot} = \mathbf{0}$ can be applied. According
to this assumption, $F_{i \cdot} = \mathbf{0}$ for most rows
$i$.

Different versions of the factor model will be explored in our
analysis. These versions differ in terms of prior formulations for
$\alpha_{il}$ and $F_{i \cdot}$. In all cases, we set the
specifications $\sigma_i^2 \sim \mathit{IG}(a,b)$ and $\lambda_{\cdot j} \sim
N_L(\mathbf{0}, I_L)$. Consider $\alpha_{il} \sim(1-h_{il}) \delta
_0(\alpha_{il}) + h_{il} N(0, \omega)$ where $h_{il}$ is a binary
indicator variable. We explore two different forms of expressing our
prior uncertainty for the probability that $h_{il} = 1$:
\begin{longlist}[(2)]
\item[(1)] $h_{il} \sim\operatorname{Bernoulli}(q_{il})$ and $q_{il} \sim
\operatorname{Beta}(\gamma_1, \gamma_2)$;
\item[(2)] $h_{il} \sim\operatorname{Bernoulli}(q_R)$, $R \in\{R_1, R_2, R_3\}$,
and $q_R \sim\operatorname{Beta}(\gamma_{1,R}, \gamma_{2,R})$. Let $R = R_1$
if we suspect that feature $i$ and factor $l$ are associated, $R = R_2$
if no association is expected, and $R = R_3$ if the relationship is
unknown.
\end{longlist}
According to specification (1), $q_{il}$ is updated using a
single observation $h_{il}$, and this strategy can be useful in
applications involving large data sets. In specification~(2), $q_{R}$ is
updated based on the group of $h_{il}$ such that $(i,l) \in R$. If the
group of indices $R_3$ contains a large number of elements and $\alpha
_{il} \neq0$ for most $(i,l) \in R_3$, the probability $q_{R_3}$ tends
to be large which favors $h_{il} = 1$. As a result, very few or none of
the $\alpha_{il}$ related to $R_3$ will be zero, that is, the level of
sparsity is lower than it should be. If $m$ is small, the model
performs well with both specifications for~$h_{il}$; see Section D in
\citet{MayrinkLucas2012} which presents a simulated study to evaluate
the performance of the models proposed in this section.

Assume a mixture prior with two components for the interaction effect
vector~$F_{i \cdot}$. One of the components is the degenerated
distribution at 0, which allows for the possibility of having $F_{i
\cdot} = \mathbf{0}$, that is, no interaction effect for feature $i$.
We will explore two versions of this mixture distribution. The first
one assumes that $F_{i \cdot}$ can be different comparing affected
features, whereas the second version assumes that $F_{i \cdot}$ is the
same for all affected features. In the context of gene expression
analysis (feature${}={}$gene), version 2 would be less realistic:
\begin{longlist}[(2)]
\item[(1)] $(F_{i \cdot}' \mid\lambda) \sim(1-z_i) \delta_0(F_{i \cdot}')
+ z_i N_n[\mathbf{0}, K(\lambda)]$,
\item[(2)] $(F_{i \cdot}' \mid F^*) \sim(1-z_i) \delta_0(F_{i \cdot}') +
z_i \delta_{F^*}(F_{i \cdot}')$ and
$(F^* \mid\lambda) \sim N_n[\mathbf{0},K(\lambda)]$,
\end{longlist}
where $z_i$ is an indicator variable and $K(\lambda)$ is
the covariance matrix obtained from the Squared Exponential covariance
function depending on $\lambda$,
%
\begin{equation}\label{Eq05}
K(\lambda)_{j_1, j_2} = \exp\biggl\{ -\frac{1}{2 l_s^2} \|
\lambda_{\cdot j_1} - \lambda_{\cdot j_2}\|^2 \biggr\},
\end{equation}
where $(j_1, j_2) \in\{1,2,\ldots,n\}$, $l_s$ is the
characteristic length-scale and $\|\mathbf{y}\|$ represents the
Euclidean norm of the vector $\mathbf{y}$. The covariance function is a
crucial ingredient in the model, as it encodes our assumptions about
the function we wish to learn. The Squared Exponential is stationary,
isotropic and probably the most widely-used kernel in the literature.
Furthermore, it is infinitely differentiable, which means that a
Gaussian Process with this choice has mean square derivatives of all
orders, and is thus smooth; see \citet{RasmussenWilliams2006}. Note
that if the points $\lambda_{\cdot j_1}$ and $\lambda_{\cdot j_2}$ are
very close in the $\mathbb{R}^L$ space, then the samples $j_1$ and
$j_2$ are similar and $K(\lambda)_{j_1, j_2} \approx1$. Conversely,
the larger the distance between these points, the higher is the
dissimilarity between the samples and the closer to 0 is $K(\lambda
)_{j_1, j_2}$. The length-scale $l_s$ is an adjustable parameter that
controls how close the points $\lambda_{\cdot j_1}$ and $\lambda_{\cdot
j_2}$ should be in order to be considered associated with each other.

We explore different strategies to express our prior knowledge about
the indicator $z_i$. Assume the following possibilities:
\begin{longlist}[(2)]
\item[(1)] $z_i \sim\operatorname{Bernoulli}(\rho_i)$ and $\rho_i \sim
\operatorname{Beta}(\beta_1, \beta_2)$;
\item[(2)] $z_i \sim\operatorname{Bernoulli}(\rho)$ and $\rho\sim\operatorname{Beta}(\beta_1,
\beta_2)$;
\item[(3)] $z_i \sim\operatorname{Bernoulli}(\rho_R)$, $R \in\{R_1, R_2\}$ and
$\rho_R \sim\operatorname{Beta}(\beta_{1,R}, \beta_{2,R})$. Here, $R = R_1$ if
we believe that feature $i$ is associated with some factor and is not
affected by interactions. Let $R = R_2$ if the association between
feature $i$ and any factor is unknown (interaction effect may exist).
\end{longlist}
Strategy (1) can be more convenient for applications
involving large $m$, because it is less influenced by other
observations. Strategy (2) assumes a global probability $\rho$
representing the level of features affected by interactions. The
updating distribution of $\rho$ takes into account all observations
$z_i$. We expect few rows of $F$ indicating nonzero effects, therefore,
$\rho$ tends to be very small if $m$ is large. This situation favors
$z_i = 0$ and, thus, the sparsity level in $F$ can be higher than
expected. This same problem can occur with $\rho_{R_2}$ in
specification (3); $\rho_R$ is updated with \mbox{$z_i\ \forall i \in R$}.

We use the structure of the Gibbs Sampling algorithm to sample from the
target distribution $p(\alpha, \lambda, F, \sigma^2 | X)$; the complete
conditional posterior distributions are presented in Section C of
\citet
{MayrinkLucas2012}. In particular, the full conditional of $\lambda
_{\cdot j}$ depends on which specification we use for $p(F_{i \cdot
}|\lambda)$. An indirect sampling method is required in this case; we
apply the Metropolis--Hastings algorithm with a random walk proposal
distribution.

\begin{table}
\tablewidth=171pt
\caption{Prior specifications defining different models} \label{Ta01}
\begin{tabular*}{\tablewidth}{@{\extracolsep{4in minus 4in}}lccc@{\hspace*{-1.5pt}}}
\hline
& \multicolumn{3}{c@{}}{\textbf{Prior distributions}} \\[-4pt]
& \multicolumn{3}{c@{}}{\hrulefill}\\
\textbf{Model} & $\bolds{h_{il}}$ & $\bolds{F_{i \cdot}}$ & $\bolds{z_i}$ \\
\hline
1 & 1 & 1 & 1 \\
2 & 1 & 2 & 1 \\
3 & 1 & 1 & 2 \\
4 & 1 & 2 & 2 \\
5 & 2 & 1 & 3 \\
\hline
\end{tabular*}
\end{table}

Table \ref{Ta01} provides an identification number for each
configuration of prior distributions defining a factor model. As can be
seen, we choose to investigate 5 different configurations. In
models~1, 3 and~5, we assume that the interaction effect can differ
from row to row in $F$. On the other hand, the same interaction effect
is considered for all affected features in models 2 and 4. Note that
model 5 is the only one using the specifications $h_{il} \sim
\operatorname{Bernoulli}(q_{R})$ and $z_i \sim\operatorname{Bernoulli}(\rho_{R})$. In
addition, models 3 and 4 apply the global Bernoulli probability $\rho$.

\section{Comparison between factor models with interactions}\label{sec4}

\begin{figure}

\includegraphics{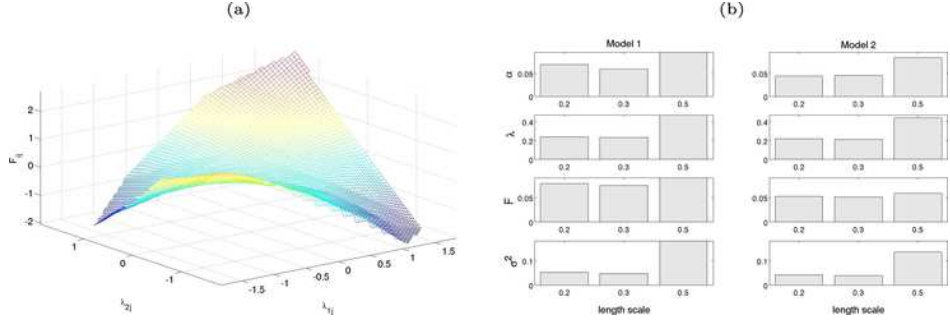}

\caption{Panel \textup{(a)}: true interaction effect in all
simulations. Panel \textup{(b)}: statistic AAD, \textup{(D.1)} in Section~\textup{D} of
Mayrink and Lucas (\citeyear{MayrinkLucas2012}), and the comparison of models 1 and 2 with
different choices of~$l_s$ (simulation 1).} \label{Fi01}
\end{figure}

Here, we compare the results from the factor models proposed in
Sections \ref{sec2} and \ref{sec3}. Consider the same data sets simulated for the
analysis in Section D of \citet{MayrinkLucas2012}. In that case, we
define $F_{ij} = \lambda_{1j} \lambda_{2j}$ as the true interaction
term affecting some features in $G_E = (G_1 \cup G_2)^C$. Figure \ref
{Fi01}(a) shows the surface plot representing the saddle shape of the
true interaction effect. Since we use the same $\lambda$ in all
simulations, this is our target interaction effect for all cases.

\begin{figure}[b]

\includegraphics{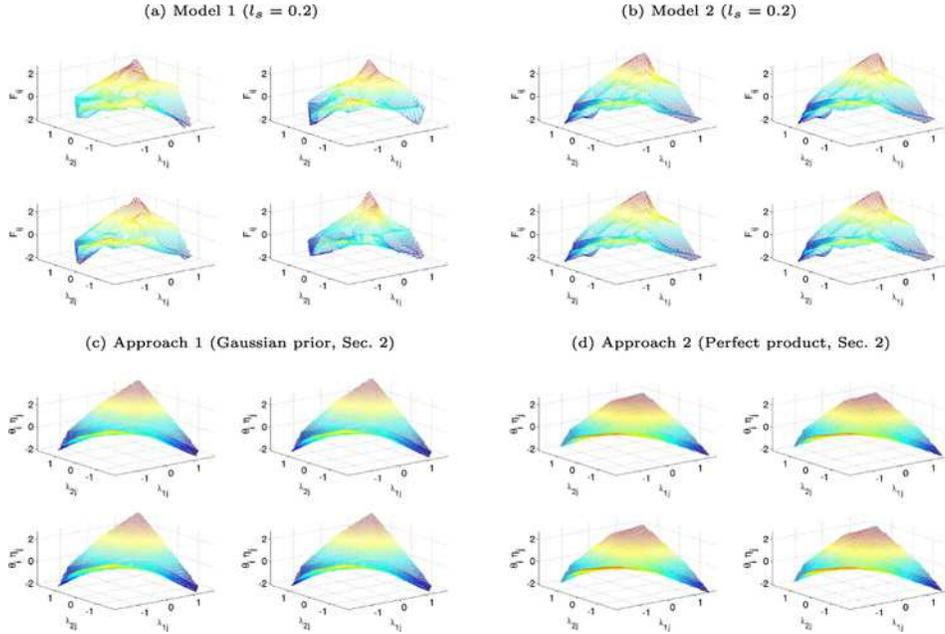}

\caption{3-D surface plot representing the estimated interaction
effect.} \label{Fi02}
\end{figure}

\begin{figure}

\includegraphics{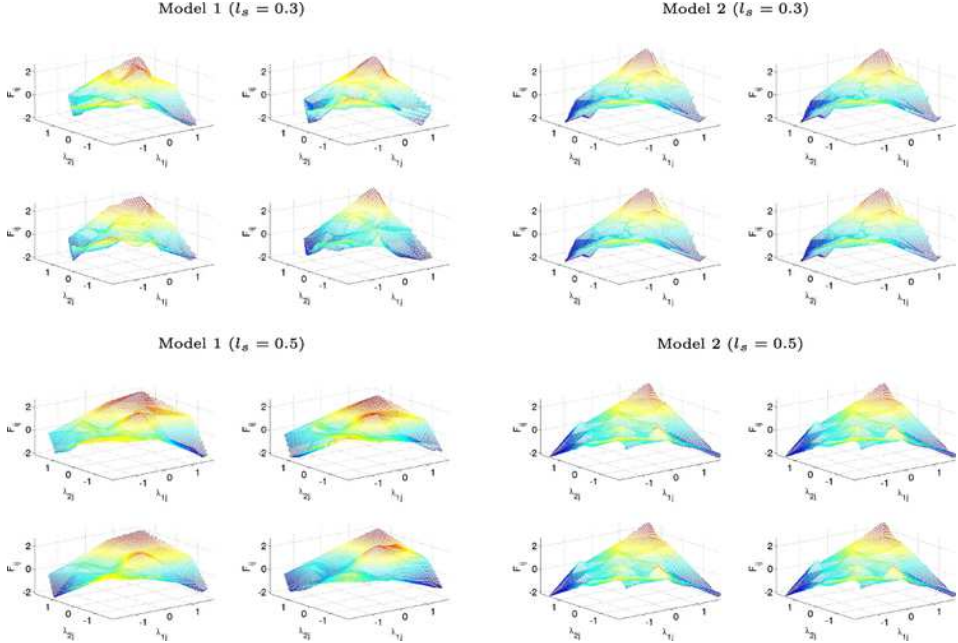}

\caption{3-D surface plot representing the estimated interaction effect
($l_s = 0.3$ or 0.5).} \label{Fi03}
\end{figure}

The model with multiplicative interactions (\ref{Eq01}) can be compared
with model (\ref{Eq04}) in Section \ref{sec3}. The interaction effect $\theta_{i
\cdot} \eta$ corresponds to $F_{i \cdot}$. Note that $\theta_{i \cdot}
= \mathbf{0}$ represents $F_{i \cdot} = \mathbf{0}$. In terms of prior
specifications, initial values and MCMC configuration, consider the
same choices defined in the simulated studies developed in Sections B
and D of \citet{MayrinkLucas2012}. In this section, we concentrate on
the comparison of surface plots to see how well the saddle shape in
Figure \ref{Fi01} is estimated.\setcounter{footnote}{2}\footnote{In order to test whether
gene $i$ is affected by interactions, we consider the conditional
probability $p(z_i = 1|\ldots)$ related to the mixture posterior
distribution of $\theta_i$ or $F_{i\cdot}$, depending on the model. If
$p(z_i = 1|\ldots) > 0.5$, we will assume a significant interaction
effect.} Figure \ref{Fi02} shows the surfaces indicating the estimated
interaction effect; we can identify the saddle shape in all cases. As
one might expect, the multiplicative model [panels~(c) and~(d)] produces a
smoother surface than the nonlinear model [panels (a) and~(b)]. The
multiplicative model is in advantage, because it assumes the true
saddle shape as the target effect. The parameter $l_s$ can be used to
control the smoothness of the surface in the nonlinear model (current
choice $l_s = 0.2$). If this value is increased, the number of
neighbors influencing each point increases; the covariance matrix is
then more populated. Figure \ref{Fi03} presents the surfaces related to
models 1 and 2 assuming bigger choices of $l_s$. As can be seen, the
level of irregularities in the middle of the graph seems reduced with
respect to $l_s = 0.2$; this conclusion is more evident for model 1
with $l_s = 0.5$.

The smooth surfaces, for $l_s = 0.5$ in Figure \ref{Fi03}, seem to be
flatter and wider than the other cases. This characteristic can be
interpreted as an indication of worse approximation between posterior
estimates and true values. The bar plots in Figure~\ref{Fi01}(b)
compare the AAD statistic, (D.1) in Section D of \citet
{MayrinkLucas2012}, for parameters in models 1 and 2 with different
choices of $l_s$. Note that the approximation is indeed worse when $l_s
= 0.5$; the biggest AAD value is observed for $l_s = 0.5$ in all cases.

Applications involving other data sets (simulations 2 and 3) and other
models (models 3, 4 and 5) provide the same conclusions above.

\section{Real application: CNA and multiplicative interactions}\label{sec5}

The number of copies of a gene in a chromosome can be modified as a
consequence of problems during cell division and these alterations are
known to play an important role in human cancer. We wish to examine the
possibility that there are genes that are synergistically affected by
copy number alteration in multiple genomic locations. In order to
assess this, we will build factor models in which we seed each latent
factor with a set of genes that is known to be in a single region of
copy number alteration (CNA). We accomplish the seeding with the prior
assumption that they have nonzero factor loadings on the factor with
very high probability. We then utilize our interaction model to assess
all genes for interaction effects between two copy number alteration
factors. Positive results will indicate genes that are synergistically
differentially expressed in the presence of multiple CNAs and may lead
to insights about the mechanism of action of the CNAs.

Many studies have detected CNA in breast cancer data, for example,
\citet
{PollackEtAl2002}, \citet{PrzybytkowskiEtAl2011} and \citet
{LucasEtAl2010}. In our analyses, different regions of CNA are drawn
from \citet{LucasEtAl2010}. Each region is an interval, involving a
collection of genes, located in the human genome sequence. The
locations suggesting CNA are known, and an annotation file identifying
the chromosome position for each probe set can be obtained from the
Affymetrix website. In order to identify our seed genes, we consider a
range (2,000,000 to the left and right) around the central
position\footnote{In \citet{LucasEtAl2010} the expression scores
of 56 latent factors were assessed on both the breast cancer data set
as well as breast tumor cell lines. These scores were then compared
with CGH clones in the corresponding tumor and cell line samples using
Pearson correlation. Approximately, $1/3$ of the factors show a
significant degree of association with the CGH clones in small
chromosomal regions in both tumor and cell line. The mentioned
``central position'' represents the central point of the chromosomal
region where the indicated correlations are significant. The analyst is
free to apply the factor model to evaluate interactions together with
any method for identification of genome regions with CNA.} where the
CNA seems to occur. We explore four different breast cancer data sets:
\citet{ChinEtAl2006}, \citet{MillerEtAl2005}, \citet
{SotiriouEtAl2006} and \citet{WangEtAl2005}.

We investigate the results for two groups of over-expressed genes. The
first one has central position 35,152,961 in chromosome 22; we denote
this group as $G_1$. The second collection of genes is located around
the central point 68,771,985 in chromosome 16; let $G_2$ represent this
group. We will fit a factor model with $L = 2$ latent factors
describing the expression pattern of the genes in $G_1$ and $G_2$. The
model includes a third factor representing the multiplicative
interaction between the first two. Our goal is to identify the genes
affected by the interaction factor.

The group $G_1$ has 50 genes, and $G_2$ contains 42 elements. As
described above, the selection of these genes is based on an interval
specified around a position in the genome. This strategy can lead to
the inclusion of cases unrelated to the CNA detected for the studied
region. In order to remove the unrelated cases from the current gene
lists, we fit a two-factor model (without interaction terms) to the
($92 \times n$) matrix $X$. The following configuration is expected for
the estimated $\alpha\dvtx\{\alpha_{i1}\dvtx  i \in G_1\}$ with the same
sign, $\{\alpha_{i2}\dvtx  i \in G_2\}$ with the same sign, and $\alpha_{il}
= 0$ for all other cases. The genes in $(G_1 \cup G_2)$ violating this
assumption are considered problematic, and thus removed from the
analysis. This cleaning procedure involving $G_1$ and $G_2$ is
described with more details in Section E of \citet{MayrinkLucas2012}.
The procedure defines 22 genes in $G_1$ and 18 in $G_2$.

Let $G_E$ represent a group of extra genes to be included in the
analysis; $G_1$, $G_2$ and $G_E$ are disjoint sets. The microarrays
selected for this application have 22,283 genes, and each breast cancer
data set has more than 100 samples available for analysis. As a result,
the MCMC algorithm can be rather slow to handle this large amount of
data. As an alternative to reduce the computational cost, we implement
a gene selection procedure to eliminate the cases which might not be
affected by interactions. The full description of the selection process
is given in Section E of \citet{MayrinkLucas2012}. In short, we
fit a
two-factor model (without interaction terms) to the ($22\mbox{,}283 \times n$)
matrix $X$ assuming 22 genes in $G_1$, 18 genes in $G_2$ and 22,243
genes in $G_E$. The distribution of the conditional probability
$p(h_{il}=1|\cdots)$ is evaluated to accept or reject $\alpha_{il} \neq
0$. It seems reasonable to assume that the genes affected by both
factors are more likely to be affected by interactions, therefore, the
final result includes only the cases satisfying this requirement. This
selection process yields 3704 genes in the updated $G_E$.

Consider the prior specifications:  $\omega_{\alpha} = \omega_{\theta} =
10$ in (\ref{Eq02}) and (\ref{Eq03}), $\sigma^2_i \sim \mathit{IG}(2.1,\allowbreak1.1)$.
Our goal is to fit the factor model with multiplicative interaction
effects (using approach 1${}={}$Gaussian prior) to the real data having 22
genes in $G_1$, 18 genes in $G_2$ and 3704 genes in $G_E$. Given the
large amount of genes, we need to set strong priors for $q_{il}$ to
impose our assumptions related to $G_1$ and $G_2$ and assure the
identification of the model. We use the configuration indicated as
``option 2'' in Table B.1. Degenerated priors are assumed to impose our
assumptions regarding the gene--factor relationship for the cases in
$G_1$ and $G_2$. This strategy is important to retain the CNA
interpretation of factors 1 and 2; otherwise, the target association
can be overwhelmed by the large amount of information in $G_E$. Note
that we assume no interaction affecting the genes in ($G_1 \cup G_2$).
The Beta($1,10$) is specified to induce sparsity in the loadings ($i
\in G_E$) related to the interaction factor. Finally, the $U(0,1)$ is
indicated for all other cases.

The MCMC algorithm performs 600 iterations (burn-in period${}={}$400). In
terms of initial values of the chains, consider the same choices
defined in Section B of \citet{MayrinkLucas2012} for $\alpha
_{il}^{(0)}$, $\lambda_{lj}^{(0)}$, $\theta_i^{(0)}$, $\eta_j^{(0)}$
and $(\sigma^2_i)^{(0)}$. The probabilities $q_{il}$ and $\rho_i$ are
initialized with the values presented in Table B.1 (option 2);
$h_{il}^{(0)} \sim\operatorname{Bernoulli}(q_{il}^{(0)})$ and $z_i^{(0)} \sim
\operatorname{Bernoulli}(\rho_i^{(0)})$. The chains seem to converge in all
applications of the MCMC algorithm.

\begin{figure}

\includegraphics{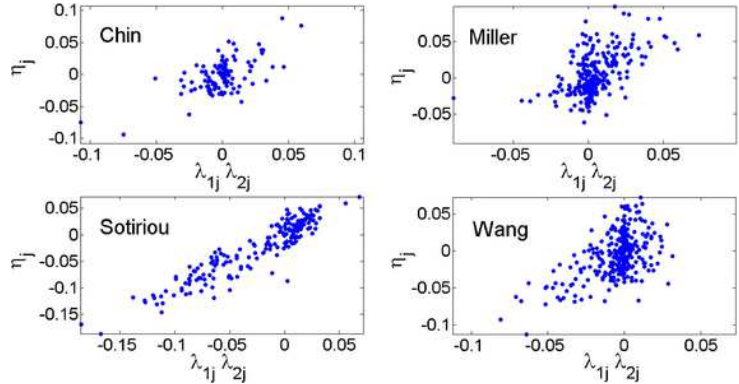}

\caption{Scatter plots comparing the posterior estimates of $\eta_j$
and $\lambda_{1j} \lambda_{2j}$ (approach 1${}={}$Gaussian prior). Each
panel represents a different breast cancer data set.} \label{Fi04}
\end{figure}

The model assuming the prior $\eta_j \sim N(\lambda_{1j} \lambda_{2j},
\nu)$ (approach 1) is the focus of the first application in the current
section. As previously discussed, the variance parameter $\nu$ must be
small to guarantee the target multiplicative effect. The real data set
contains a large number of genes and, thus, the posterior variance is
expected to be small. In this case, only extremely small values for $\nu
$ will ensure that $\eta_j$ and $\lambda_{1j} \lambda_{2j}$ are
correlated. Figure \ref{Fi04} shows scatter plots comparing the
posterior estimates of $\eta_j$ and the product $\lambda_{1j} \lambda
_{2j}$. Here, the factor model is fitted with $\nu= 10^{-5}$. Note
that the model fit for the data set ``Sotiriou'' is the only one
indicating correlated results. In the other applications, the
multiplicative effect is lost and the interaction factor is just
another factor.

\begin{figure}

\includegraphics{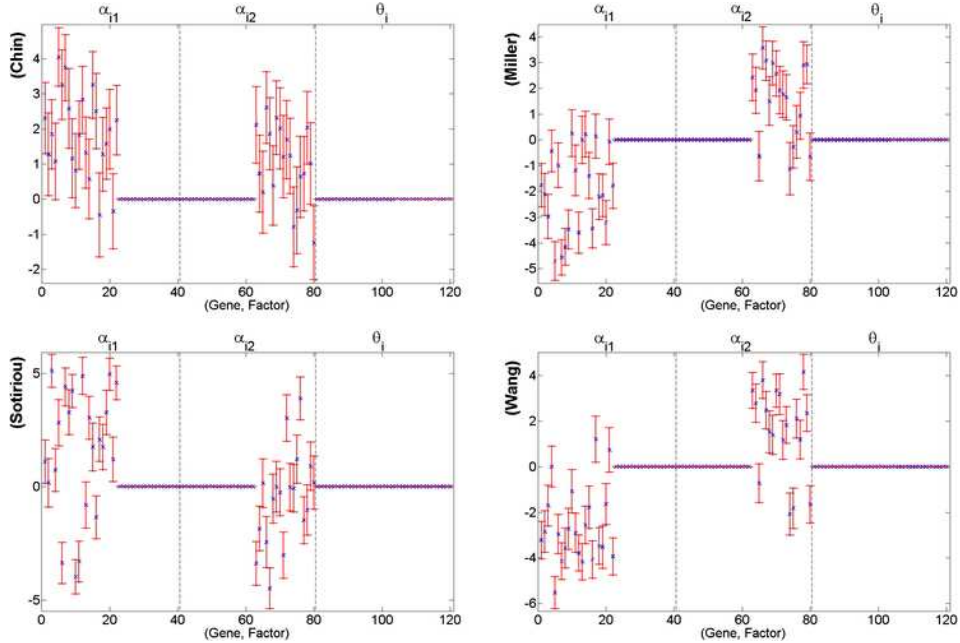}

\caption{Posterior mean (x mark) and the 95\% credible interval (bar)
for the loadings with $i \in(G_1 \cup G_2)$ (approach 2${}={}$perfect
product). Intervals are computed for the component with highest
posterior probability weight. Dashed lines separate the factors.}
\label{Fi05}
\end{figure}

Given the difficulty to set $\nu$, no further real data analysis is
developed for the factor model with approach 1. Our next step is to
investigate the model defined as approach 2, where we force the perfect
association $\eta_j = \lambda_{1j} \lambda_{2j}$. Consider the same
breast cancer data sets, configuration of prior distributions, initial
values and MCMC setup defined in the previous application. Because we
impose the equality between $\eta_j$ and $\lambda_{1j} \lambda_{2j}$,
the scatter plots comparing their values indicate correlation 1. Figure
\ref{Fi05} shows the 95\% credible interval and the posterior mean for
$\alpha_{il}$ and $\theta_i$ such that $i \in(G_1 \cup G_2)$. Note
that most nonzero loadings, related to the same factor, indicate
posterior estimates with the same sign. This fact is observed for all
data sets, and it supports the CNA interpretation for factors 1 and 2.
Recall that the zero estimates are imposed via prior distribution to
satisfy our assumptions for this group of genes.

Table \ref{Ta02} indicates (main diagonal) the number of genes affected
by multiplicative interactions in each real data application. Note that
the majority of features are free from interaction effects. The
elements off diagonal are the number of common genes belonging to the
intersection between the groups of affected genes. As can be seen, at
least 14 genes can be found in the intersections involving different
data sets. This result may be used as an argument against the idea that
the model might be identifying interactions for a random set of genes.
The intersections involving three data sets have 2--6 elements. Only 1
gene belongs to the intersection of all four data sets; its official
full name is ``GTP binding protein 4,'' and it is located in chromosome~10.

%
\begin{table}[b]
\tablewidth=200pt
\caption{Pairwise intersections between data sets; number of common
genes affected by the multiplicative interaction}
\label{Ta02}
\begin{tabular*}{\tablewidth}{@{\extracolsep{\fill}}lcccc@{}}
\hline
& \textbf{Chin} & \textbf{Miller} & \textbf{Sotiriou} & \textbf{Wang} \\
\hline
Chin & 314 & \hphantom{0}30 & \hphantom{0}24 & \hphantom{0}20 \\
Miller & \hphantom{0}30 & 170 & \hphantom{0}14 & \hphantom{0}24 \\
Sotiriou & \hphantom{0}24 & \hphantom{0}14 & 244 & \hphantom{0}24 \\
Wang & \hphantom{0}20 & \hphantom{0}24 & \hphantom{0}24 & 255 \\
\hline
\end{tabular*}
\end{table}

We apply a hypothesis test to investigate whether the configuration in
Table \ref{Ta02} can be considered a result of an independent random
sample of genes, from the population of 3704 cases in $G_E$, for each
breast cancer data set. First, we select genes, uniformly at random,
using the numbers in the main diagonal of Table \ref{Ta02} as the
sample sizes. In the next step, we consider the pairwise intersections
between the random selections and obtain the sum of elements in all
intersections; this number $n_k$ represents the level of overlaps. We
repeat this procedure 100,000 times to generate $\{ n_k\dvtx  k = 1, 2,
\ldots, 100\mbox{,}000\}$. Finally, we calculate the number of cases such
that $n_k \geq n_{\mathrm{o}}$, where $n_{\mathrm{o}}$ is the overlap level observed in Table
\ref{Ta02}. This result is then divided by 100,000 to provide the
$p$-value 0.00003. In conclusion, we reject the hypothesis that the genes
are independently selected for each data set.

\begin{figure}

\includegraphics{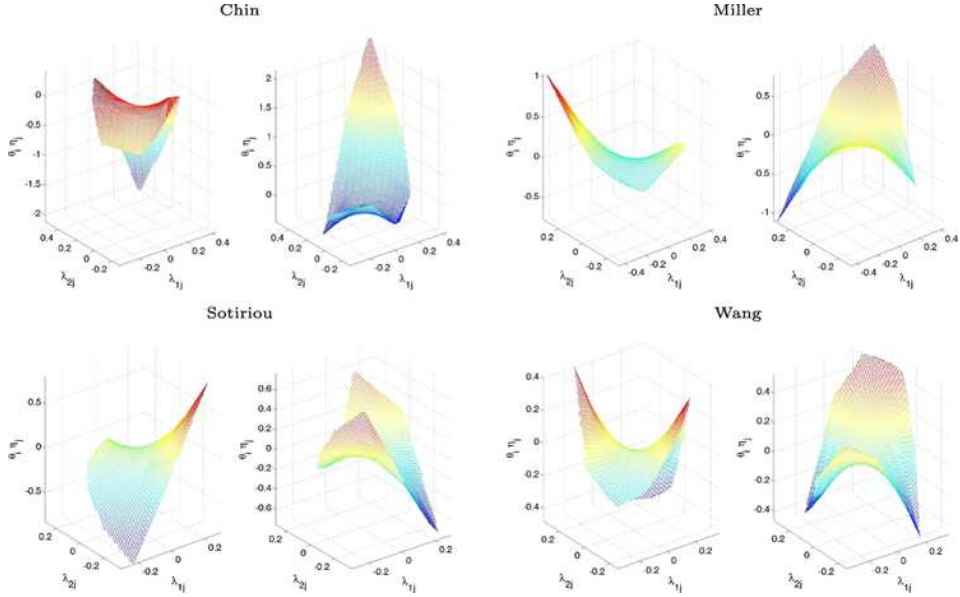}

\caption{3-D surface plots representing the multiplicative interaction
effect $\theta_i \eta_j$ (approach 2${}={}$perfect product). In each panel,
left${}={}$the smallest negative loading, and right${}={}$the largest positive
loading.} \label{Fi06}
\end{figure}

Figure \ref{Fi06} shows the three-dimensional surface plot representing the
multiplicative effect associated with the genes with the highest
interaction effects. As can be seen, this type of interaction has a
saddle shape. Each point in the surface corresponds to a different
sample $j$. In the x and y axes we have $\lambda_{1j}$ and $\lambda
_{2j}$; the z axis represents $\theta_i \eta_j$. The loading $\theta_i$
controls how strong the interaction effect is; values close to zero
define flatter surfaces. The sign of $\theta_i$ determines the
orientation of the saddle. In each panel, the graph on the left is
related to the smallest negative~$\theta_i$, while the graph on the
right represents the largest positive $\theta_i$.

\section{Real application: CNA and nonlinear interactions}\label{sec6}

Consider again the CNA problem investigated in the previous section
using the four breast cancer data sets: \citet{ChinEtAl2006},
\citet
{MillerEtAl2005}, \citet{SotiriouEtAl2006} and \citet
{WangEtAl2005}. Two latent factors are defined in our model for this
type of application. In other words, $\lambda$ has two rows of factor
scores, and each row describes the expression pattern across samples
for the genes associated with a region where the CNA was detected. We
will evaluate the model fit assuming three different pairs of
chromosome locations. Table \ref{Ta03} identifies the position and
chromosome number for each region. Denote by $G_1$ the group of genes
around the first location in the pair; $G_2$ represents the collection
of features around the second location. The cleaning procedure,
described in Section E of \citet{MayrinkLucas2012}, is applied to
remove problematic genes from $G_1$ and $G_2$. Table \ref{Ta03}
indicates the number of genes before and after the removal procedure.

\begin{table}
\tablewidth=270pt
\caption{Regions detected with CNA. We apply a procedure to remove
genes unrelated to the CNA factors. The number of genes before and
after this removal is presented} \label{Ta03}
\begin{tabular*}{\tablewidth}{@{\extracolsep{4in minus 4in}}lcccc@{}}
\hline
& & & \multicolumn{2}{c@{}}{\textbf{Number of genes}} \\[-4pt]
& & & \multicolumn{2}{c@{}}{\hrulefill}\\
\textbf{Region} & \textbf{Chr.} & \textbf{Position} & \textbf{Before}
& \textbf{After} \\
\hline
1 & 11 & 117,844,879 & 38 & 13 \\
2 & 22 & \hphantom{0}35,152,961 & 50 & 22 \\
3 & \hphantom{0}7 & 101,400,207 & 45 & 24 \\
4 & 16 & \hphantom{0}68,771,985 & 42 & 18 \\
\hline
\end{tabular*}
\end{table}

The microarrays have 22,283 genes and each data set contains at least
118 samples. In order to reduce the computational cost, consider again
the gene selection procedure described in Section E of \citet
{MayrinkLucas2012}. The method is based on the data set in \citet
{ChinEtAl2006}, and we evaluate the pairs of regions $(1,4)$, $(2,4)$ and
$(3,4)$; see Table \ref{Ta03}. The selection indicates 3717, 3704 and
3708 elements in $G_E$ for the pairs $(1,4)$, $(2,4)$ and $(3,4)$. For the
purpose of comparison, this configuration of $G_E$ is used to study all
data sets. Our goal is to identify features in $G_E$ affected by interactions.

Model 1 in Table \ref{Ta01} is more convenient for applications
with large $m$. In this case, we assume a particular Bernoulli
probability for each indicator $h_{il}$ and~$z_i$, which makes these
variables less dependent on other observations. If a large number of
$h_{il}$ share the same Bernoulli probability $q_R$, the level of
sparsity in $\alpha$ can be incorrectly determined. If most loadings
are nonzero values, $q_R$ tend to be large which favors $h_{il} = 1$
for all $(i,l)$ related to $q_R$. Similarly, if a large number of $z_i$
share the same probability $\rho$ (models 3, 4) or $\rho_R$ (model 5),
and if $F_{i \cdot} = \mathbf{0}$ for most genes, then $\rho$ or $\rho
_R$ tend to be small which favors $z_i = 0$ for all involved features.
Here, the level of sparsity is too high and some interaction effects
are neglected. In a real application, it seems more realistic to assume
different interaction effects for different affected genes; for this
reason, model 1 is preferred to model 2.

Assume $\omega= 10$ in the mixture prior for $\alpha_{il}$, $\sigma^2_i
\sim \mathit{IG}(2.1,1.1)$, and set $l_s = 0.2$ in (\ref{Eq05}). The
specifications in Table D.1 (option 2) are defined for $q_{il}$ and
$\rho_i$ to impose our assumptions regarding the gene--factor
relationship and provide the identification of the model. We do not
expect interaction effects related to the genes in $G_1$ and $G_2$;
these groups have a strong relationship with one latent factor and no
association with the other. In addition, recall that most rows of $F$
should be null-vectors to ensure the identification between $\alpha
\lambda$ and $F$. It is reasonable to expect few genes affected by
interactions; as a result, one might choose a Beta distribution with
higher probability mass below 0.5 for $\rho_i$ with $i \in G_E$. The
choice $\rho_i \sim\operatorname{Beta}(1,1)$ works well in the applications of
this section.

In terms of initial values of the chains, let $F_{ij}^{(0)} = 0$ for
all $(i,j)$, and consider the usual choices $\alpha_{il}^{(0)} = 0$,
$(\sigma^2_i)^{(0)} = 1$, and $\lambda_{lj}^{(0)} \sim N(0,1)$. We
initialize $h_{il}^{(0)} \sim\operatorname{Bernoulli}(q_{il}^{(0)})$ and
$z_i^{(0)} \sim\operatorname{Bernoulli}(\rho_i^{(0)})$, where $q_{il}^{(0)}$
and $\rho_i^{(0)}$ are indicated in Table D.1 (option 2). The MCMC
algorithm is set to perform 600 iterations (burn-in period${}={}$300); the
chains seem to converge in all applications. The Metropolis--Hastings
algorithm, used to sample from the full conditional posterior
distribution of $\lambda_{\cdot j}$, has acceptance rate around
31--40\%, 15--65\%, 26--53\% and 67--84\% in the applications related to
the data sets [\citet{ChinEtAl2006}, \citet{MillerEtAl2005},
\citet
{SotiriouEtAl2006} and \citet{WangEtAl2005}].

\begin{figure}

\includegraphics{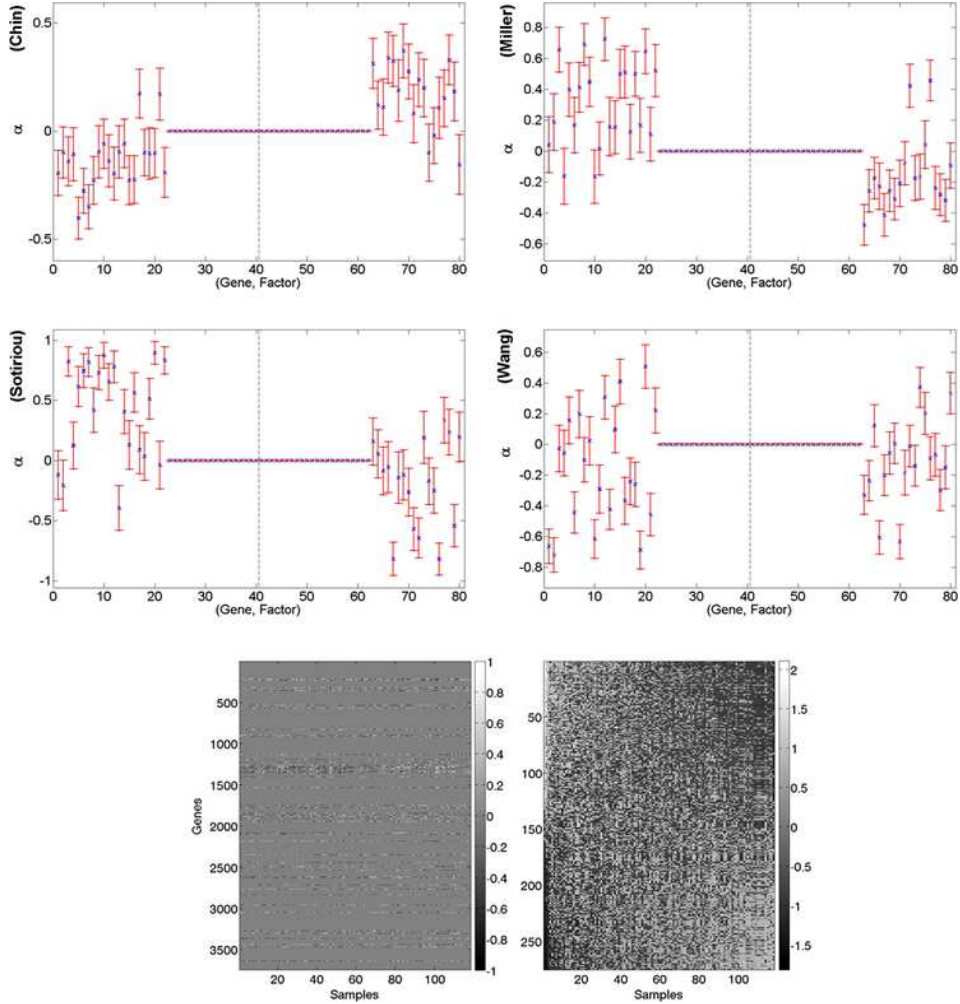}

\caption{Results related to the pair of locations $(2,4)$. First four
panels: posterior mean (x mark) and 95\% credible interval (bar) for
$\alpha_{il}$ with $i \in(G_1 \cup G_2)$; the dashed line separates
the two factors. Fifth panel: left-hand side${}={}$full matrix $F$ (3744 genes),
right-hand side${}={}$cases $F_{i \cdot} \neq0$ (rows and columns are
sorted so
that the 1st principal components are monotone).} \label{Fi07}
\end{figure}

The 5th panel in Figure \ref{Fi07} shows images of interaction
effects in $F$. The image on the left represents the full matrix with
3744 rows and 118 columns; the color bar is constrained between $(-1,1)$
for higher contrast. The second heat map exhibits the cases $F_{i \cdot
} \neq0$. Note that we identify 275 genes affected by nonlinear
interactions involving the factors. Further, the second image suggests
a coherent pattern for groups of features; several rows have similar
decreasing or increasing effect, as we move across samples. This result
supports the idea of $F_{i \cdot}$ as a representation of interactions;
on the contrary, a random pattern would be observed for most rows.
Figure \ref{Fi07} also presents the posterior estimates and 95\%
credible interval for the loadings related to genes in $G_1$ and $G_2$.
These results are computed for the component in the posterior mixture
with the highest probability weight. As can be seen, most intervals in
$G_l$, $l = 1$ or 2, suggest loadings with the same sign. This result
supports the association between factors 1--2 and the CNA detected for
$G_1$ and $G_2$. In other words, the estimated interactions seem to be
a result of the CNA in regions 2 and 4.

\begin{figure}

\includegraphics{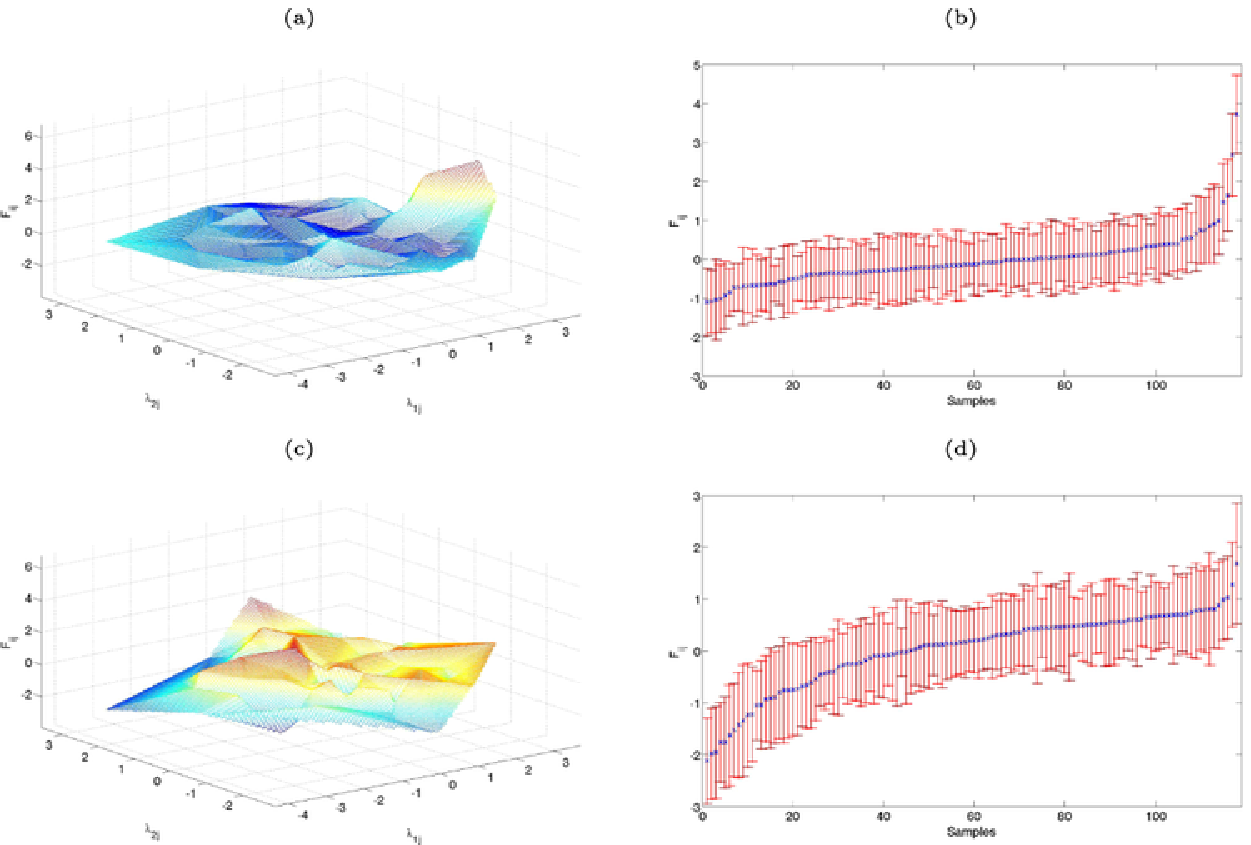}

\caption{3-D surface plot of the estimated interaction effect $F_{1524
\cdot}$ \textup{(a)} and $F_{1945 \cdot}$ \textup{(c)}. Panels \textup
{(b)} and \textup{(d)} contain the
posterior mean (x mark) and the 95\% credible interval (bar). This
result is related to the data set Chin et~al. (\citeyear{ChinEtAl2006}) and the
pair of
locations~$(2,4)$.} \label{Fi08}
\end{figure}

Figure \ref{Fi08} shows, in panels (a) and (c), the three-dimensional
surface plot representing the shape of the estimated interaction effect
for two genes. The x and y axes contain the estimated $\lambda_{1j}$
and $\lambda_{2j}$, therefore, each point in the x--y plane is related
to a sample (microarray). These shapes are different, suggesting
distinct interaction effects for those genes. Panels (b) and (d)
present the posterior mean used in the z axis of the graph and the
corresponding 95\% credible interval indicating our posterior
uncertainty related to the estimated surface.

Table \ref{Ta04} compares the list of affected genes related to
different breast cancer data sets. The table is divided in three
sections representing the pair of regions with CNA. The main diagonal
in each section indicates the number of affected genes. Note that all
intersections are nonempty sets, that is, different data sets indicate
the same group of genes as affected by interactions. Given the large
number of genes in $G_E$ and the relatively small list of affected
cases determined in each application, the identification of elements in
the intersections is an important result suggesting a plausible model.
Most intersections involving three data sets have 1 or 2 elements for
any pair of regions.

\begin{table}
\tablewidth=259pt
\caption{Intersections between data sets; common genes affected by
interactions} \label{Ta04}
\begin{tabular*}{\tablewidth}{@{\extracolsep{\fill}}lcccc@{}}
\hline
& \textbf{Chin}
& \textbf{Miller} & \textbf{Sotiriou} & \textbf{Wang} \\
\hline
\multicolumn{5}{@{}c@{}}{Pair $(1,4)$}\\[4pt]
Chin & 139 & \hphantom{00}6 & \hphantom{00}8 & \hphantom{00}9 \\
Miller & \hphantom{00}6 & \hphantom{0}81 & \hphantom{00}6 & \hphantom{00}3 \\
Sotiriou & \hphantom{00}8 & \hphantom{00}6 & 121 & \hphantom{00}1 \\
Wang & \hphantom{00}9 & \hphantom{00}3 & \hphantom{00}1 & \hphantom{0}46 \\
[4pt]
\multicolumn{5}{@{}c@{}}{Pair $(2,4)$}\\[4pt]
Chin & 275 & \hphantom{0}14 & \hphantom{0}13 & \hphantom{0}19 \\
Miller & \hphantom{0}14 & 111 & \hphantom{00}7 & \hphantom{00}7 \\
Sotiriou & \hphantom{0}13 & \hphantom{00}7 & 143 & \hphantom{00}8 \\
Wang & \hphantom{0}19 & \hphantom{00}7 & \hphantom{00}8 & 111 \\
[4pt]
\multicolumn{5}{@{}c@{}}{Pair $(3,4)$}\\[4pt]
Chin & 235 & \hphantom{0}10 & \hphantom{0}11 & \hphantom{00}7 \\
Miller & \hphantom{0}10 & \hphantom{0}91 & \hphantom{00}4 & \hphantom{00}9 \\
Sotiriou & \hphantom{0}11 & \hphantom{00}4 & 115 & \hphantom{00}2 \\
Wang & \hphantom{00}7 & \hphantom{00}9 & \hphantom{00}2 & \hphantom{0}75 \\
\hline
\end{tabular*}
\end{table}

We evaluate the results of Table \ref{Ta04} to test the hypothesis of
independent random samples of genes for each data set. This same test
was used in Section \ref{sec5} to examine Table \ref{Ta02}. The configuration of
Table \ref{Ta04} provides the $p$-values: 0.00002 for the pair $(1,4)$,
0.00001 for $(2,4)$ and 0.00044 for $(3,4)$. Assuming a significance level
of 0.05, we reject the indicated null hypothesis.

In our final comparison analysis, the frameworks approach 1 (Section \ref{sec2})
and model 1 (Section \ref{sec3}) have been used to fit the data sets [\citet
{ChinEtAl2006}, \citet{MillerEtAl2005}, \citet{SotiriouEtAl2006}
and \citet{WangEtAl2005}]; consider the pair of regions $(2,4)$ in Table
\ref{Ta03}. Each model provides a list of genes affected by
interactions; we have found 22 (Chin), 7 (Miller), 13 (Sotiriou) and 7
(Wang) genes in the intersection of the lists generated for the same
data set. This type of result reinforces the idea that the proposed
models can be valid to study interactions.

\section{Conclusions}\label{sec7}

In an ordinary factor analysis, the involvement of any feature with the
factors is always additive. Biological pathways establishing complex
structure of dependencies between genes motivate the idea of a
multi-factor model with interaction terms. We study the expression
pattern across samples using Affymetrix GeneChip\regtm\ microarrays. The
matrix $X$ contains the preprocessed data (RMA outputs) with rows
representing genes and columns representing microarrays. Each column is
a different individual, but all samples are related to the same type of
cancer cell. We formulate the factor models with spike and slab prior
distributions to allow for sparsity and then test whether the effect of
factors/interactions on the features is significant or not. Simulated
studies have been developed to verify the performance of the proposed
models; the posterior estimates approximate well the real values.

In Section \ref{sec2} we have proposed a model with pairwise multiplicative
interactions, but any function defining a relationship between a pair
of factors can be used. Two approaches were considered to introduce the
interaction effect: (1) the product is inserted as the mean of a
Gaussian prior, (2) we assume the perfect product between factors in a
deterministic setup. In the real data application we have studied four
breast cancer data sets. Two factors were defined in the model, and
each one is directly associated with the genes located in a particular
region (detected with CNA) of the human genome. The main aim was to
identify other genes affected by the product interaction of the two
factors. A selection process was implemented to choose the most
interesting genes for this study, nevertheless, the matrix $X$
represents a large number of features. In this case, approach 1
requires a Gaussian prior with extremely small variance to ensure the
multiplicative effect. On the other hand, approach 2 does not suffer
from the same problem given its deterministic formulation. Depending on
the data set, we have observed 170--314 genes affected by interactions,
and the pairwise intersections of these groups have at least 14 elements.

In Section \ref{sec3} we have developed a multi-factor model with a nonlinear
structure of interactions; this version is more general. The
nonlinearities involving the latent factors were introduced through the
Squared Exponential kernel, which defines the covariance matrix in the
Gaussian component of a mixture prior specified for the parameter
representing interaction effects. One version of this prior assumes
that the effect can be different comparing affected genes; the less
realistic assumption ``same effect for any pair of affected features''
was also studied. In addition, different prior formulations were
considered for probability parameters in the mixture prior specified
for the interaction effects and for the factor loadings. As a result,
five versions of the model were defined for investigation. Assumptions
related to the intended type of application were used to choose the
priors and induce a specific configuration in the matrices of factors
loadings and interaction effects, which provides the identification of
the model. In the real data application, we have revisited the
two-factor analysis based on regions with CNA. Four breast cancer data
sets were explored, and interactions can be identified in all
evaluations. The intersections of results from the four data sets are
nonempty sets which suggest a plausible model.

The use of a different covariance function can be an alternative to
better combine smoothness and good posterior estimation. Of particular
interest in this regard is the Matern class of covariance functions
$K(r) = [2^{1-\upsilon}/\Gamma(\upsilon)] (r\sqrt{2\upsilon
} /l_s)^{\upsilon} \*K_{\upsilon}(r\sqrt{2\upsilon}/l_s)$ with positive
parameters $\upsilon$ and $l_s$, where $K_\upsilon$ is a modified
Bessel function [see \citet{AbramowitzStegun1965}, Section 9.6] and
$r$ is the Euclidean length. The parameter $\upsilon$ is, in fact, a
smoothness parameter. The Squared Exponential covariance function
$\exp\{ -r^2/(2 l_s^2) \}$ is obtained for $\upsilon=
\infty$ [see \citet{RasmussenWilliams2006}, page 204]. The process is
$k$-times Mean Squared differentiable if and only if $\upsilon> k$. In
summary, we currently control the range of influence between points
using the parameter $l_s$. In order to improve smoothness and retain
good posterior approximation, one could try to balance the choices of
$l_s$ and $\upsilon< \infty$.

In Section \ref{sec3} we have studied two mixture priors for $F_{i \cdot}$
specifying extreme cases, that is, the effects are all different or the
same. It would be reasonable to consider the intermediate situation,
where we identify groups of genes such that the nonlinear interaction
is the same within each group, but it differs between groups. In order
to implement this assumption, we can use the clustering properties of
the Dirichlet Process (DP) [Ferguson (\citeyear{Ferguson1973,Ferguson1974})]. The
following result is implied by the Polya urn scheme in \citet
{BlackwellMacQueen1973}, and it leads to the so-called ``Chinese
Restaurant Process'' [see \citet{Aldous1985}, page~92]:\vspace*{1pt}
$(\psi_i |\psi_1,\ldots,\psi_{i-1}) \sim[\zeta/(\zeta+i-1)] P_0 + \sum
_{j=1}^{i-1} [1/(\zeta+i-1)] \delta_{\psi_j}$, where $\zeta$ is the
concentration parameter and $P_0$ is the base distribution in the DP.
This implies that the $i$th feature is drawn from a new cluster with
probability proportional to $\zeta$ or is allocated to an existing
cluster with probability proportional to the number of features in that
cluster. As a result, we can consider the prior $(F_{i
\cdot}' \mid\lambda) \sim(1-\rho_i) \delta_0(F_{i \cdot}) + \rho_i
DP(\zeta, P_0)$ with $P_0 = N_n[\mathbf{0},K(\lambda)]$,
where $K(\lambda)$ is the covariance matrix depending on $\lambda$.

\section*{Acknowledgments}

The authors would like to thank Mike West, Sayan Mukherjee and the
anonymous referees for constructive comments.

\begin{supplement}
\stitle{Sparse latent factor models with interactions: Posterior
computation, simulated studies and gene selection procedure\\}
\slink[doi]{10.1214/12-AOAS607SUPP} 
\sdatatype{.pdf}
\sfilename{aoas607\_supp.pdf}
\sdescription{Additional material containing the following:
formulations of the complete conditional posterior distributions for
parameters in the proposed models, simulated studies to evaluate the
performance of the models, and the description of the procedure used to
select genes for the real applications.}
\end{supplement}


\printaddresses

\end{document}